\documentclass[twocolumn]{aastex63}

\newcommand{\align}

\accepted{October 8, 2021}

\shorttitle{The Sizes and Dynamical Motions of Granules}
\shortauthors{Liu et al.}

\begin{document}

\title{Investigations of Sizes and Dynamical Motions of Solar Photospheric Granules by a Novel Granular Segmenting Algorithm}

\correspondingauthor{Jiang,Chaowei; Yuan,Ding}
\email{chaowei@hit.edu.cn;yuanding@hit.edu.cn; liuyanxiao@hit.edu.cn}
\author{Liu,Yanxiao}
\affiliation{Institute of Space Science and Applied Technology, Harbin Institute of Technology, Shenzhen, Guangdong 518055, China}
\author{Jiang, Chaowei}
\affiliation{Institute of Space Science and Applied Technology, Harbin Institute of Technology, Shenzhen, Guangdong 518055, China}
\author{Yuan, Ding}
\affiliation{Institute of Space Science and Applied Technology, Harbin Institute of Technology, Shenzhen, Guangdong 518055, China}
\author{Zuo, Pingbing}
\affiliation{Institute of Space Science and Applied Technology, Harbin Institute of Technology, Shenzhen, Guangdong 518055, China}
 \author{Wang, Yi}
 \affiliation{Institute of Space Science and Applied Technology, Harbin Institute of Technology, Shenzhen, Guangdong 518055, China}
  \author{Cao, Wenda}
 \affiliation{New Jersey Institute of Technology, Center for Solar Research, 323 Martin Luther King Blvd., Newark, NJ, USA;
 	 Big Bear Solar Observatory, 40386 North Shore Lane, Big Bear City, CA, USA}

\begin{abstract}
Granules observed in solar photosphere are believed to be convective and turbulent, but the physical picture of granular dynamical process remains unclear. Here we performed an investigation of granular dynamical motions of full length scales based on data obtained by the 1-meter New Vacuum Solar Telescope (NVST) and the 1.6-meter Goode Solar Telescope (GST). %
We developed a new granule segmenting method, which can detect both small faint and large bright granules. A large number of granules were detected and two critical sizes, 265 km and 1420 km, were found to separate the granules into three length ranges. The granules with sizes above 1420 km follow Gaussian distribution, and demonstrate ``flat'' in flatness function, which shows that they are non-intermittent and thus are dominated by convective motions. Small granules with sizes between 265 and 1420 km are fitted by a combination of power law function and Gauss function, and exhibit non-linearity in flatness function, which reveals that they are in the mixing motions of convection and turbulence. 
Mini granules with sizes below 265 km follow power law distribution and demonstrate linearity in flatness function, indicating that they are intermittent and strongly turbulent. 
 These results suggest that a cascade process occurs: large granules break down due to convective instability, which transport energy into small ones; then turbulence is induced and grows, which competes with convection and further causes the small granules to continuously split. Eventually, the motions in even smaller scales enter in a turbulence-dominated regime. 
\end{abstract}

\keywords{Sun:photosphere $-$ Sun: granule $-$ Sun: dynamical motions}


\section{Introduction}\label{sec:intro}
Granules are overshoots of convective motions from sub-photosphere, and their patterns are formed by granule lanes where cold mass dropping backward to convection zone (\citealt{1998ApJ...499..914S,1995ApJ...443..863R}).
They are considered highly dynamic and turbulent characterizing by random motions and large Reynolds numbers.
With high values of Reynolds numbers, vortices are generated due to shear instability, and energy is transported from large scale eddies and vortices to small scale ones and down to micro-scale ones. At last, they are dissipated by viscosity. This cascade process could cause the population of granule cells increase significantly with the decreasing scales, which is often described by the relationship between energy and wavenumber of $E_{k} \sim k^{-5/3}$, according to the theory of Kolmogorov spectrum (\citealt{1941DoSSR..30..301K,1991RSPSA.434....9K}) under the assumption of isotropic and homogeneous turbulence in three dimensions. 
 
 According to Kolmogorov theory, the relationship of $P \sim A^{D/2}$  between perimeter ($P$) and area ($A$) of turbulent cells is predicted  with fractal dimension $D$ of $\frac{5}{3}$ for isotherms (\citealt{1977fgn..book.....M}) and  of $\frac{4}{3}$ for isobars (\citealt{1982Sci...216..185L}).
 \cite{1986SoPh..107...11R} found the fractal dimension values of 1.25 and 2.15 with granule sizes smaller and larger than 1{\arcsec}.37, respectively. 
 \cite{1997ApJ...480..406H} obtained a similar result with fractal dimension values of 1.3 and 2.1 for granule sizes smaller and larger than 1{\arcsec}.39, respectively. 
 \cite{1986SoPh..107...11R} explained that the granules with sizes smaller than 1{\arcsec}.37  might be the turbulent origin, and the granules with sizes larger than 1{\arcsec}.37 are convective. They further suggested that, if the small granules are turbulent, it is necessary to classify the granules into convective and turbulent. Namely, the large and middle ones are granule cells and the  small ones are ``photospheric turbulent elements". 
Although \cite{1991afa..conf...77B} and \cite{1990VA.....33..413G} confirmed these results, they argued that there is no obvious evidence to support the small granules being the turbulent origin since the fractal dimension value might be strongly affected by technical problems, i.e., the limited resolution, the definition of granules and so on.
To avoid the uncertainties caused by granular segmenting thresholds, telescope limited resolution as well as other possible factors, \cite{2012ApJ...756L..27A} calculated the flatness function, which is a method of measuring the intermittency and multi-fractality and is in dependent of length scales (\citealt{2010ApJ...722..122A,2005SoPh..228...29A}), and confirmed \cite{1986SoPh..107...11R} results.  
\cite{2012ApJ...756L..27A} found that the size distribution of large granules follow Gauss function and these large granules are non-intermittent. These regular granules represent ``dominant" granules with sizes larger than 1000 km (\citealt{1986SoPh..107...11R}). The mini granules follow power law distribution with index of $-1.82 \pm 0.12$, and they are considered to be highly intermittent and multi-fractal with sizes below 600 km.

Additionally, the power spectra of velocity fluctuations and intensity fluctuations, which represent the kinetic and thermal energies of granules, follow power-law functions with slopes of $-\frac{5}{3}$ and $\frac{11}{3}$, and thus reveal a turbulent cascade process of energy transportation from large granule cells to small ones (\citealt{1994A&A...285..322S,1995A&AS..109...79E}). Nevertheless, \cite{2001SSRv...95....9P} pointed out that the physical conditions dominating in the photosphere
 are strongly inhomogeneous and anisotropy, and are more complicated than the turbulent state described by Kolmogorov theory.

As such, the dynamics of fluid motions including convective and turbulent motions in the photosphere are still under debate and more detailed studies are required. 
To help understanding the granular dynamical process, we firstly developed a new granule segmenting algorithm and then detected two critical size points separating the granular dynamical motions into three regimes: convection, mixing motions of convection and turbulence, and turbulence. This paper is organized as follows: Section \ref{sec:Observation} describes the observations and data processing, Section \ref{sec:Methodology} describes the segmentation method, Section \ref{sec:Result} presents the results and discussions, and Section \ref{sec:Conclusion} gives a conclusion of this work.

\section{Observations and data processing}\label{sec:Observation}
Two data sets are analyzed in this work, and the first one was obtained by NVST at TiO (7058{\AA}) wavelength on November 27, 2019 in the Fuxian Solar Observatory (FSO, \citealt{2014RAA....14..705L}). The very good and stable seeing in FSO allows the spatial resolution of NVST to achieve 0.2{\arcsec}, which is close to the diffraction limit resolution. 
The observation was chosen in a quiet region near disk center with a field of view of 99{\arcsec} ${\times}$82{\arcsec}, a cadence of 30 s, and pixel size of 0.039{\arcsec}.
The selected data were firstly processed with dark current subtraction and flat field correction, and then reconstructed with speckle masking method (\citealt{2016NewA...49....8X}). 
At last, a high precision alignment was further applied to the reconstructed data (\citealt{2015RAA....15..569Y,6463241,2020RAA....20..103W}).

The second data set was taken by 1.6-meter GST (\citealt{2010AN....331..636C, 2012SPIE.8444E..03G}) at wavelength of TiO 7057 {\AA} on October 03, 2019 in Big Bear Solar Observatory (BBSO).  The data were obtained in a very quiet region near disk center, and they were taken under a very good seeing condition with the assistance of adaptive optics (\citealt{2014SPIE.9148E..35S,2010AN....331..636C}), which allows the spatial resolution of GST close to 0.1{\arcsec}.
The field of view is 69{\arcsec}${\times}$69{\arcsec}, and the cadence is 10 s.
Images owning very high quality and having pixel size of 0.034{\arcsec} after reconstruction (\citealt{2007ApOpt..46.8015W}). 
We selected these two data sets and try to analyze the dynamics of granules by data of different spatial resolutions.

\section{Method of segmentation and identifying granules}\label{sec:Methodology}
In order to extract the granule cells, a method including  granular segmentation and identifying is developed.
The segmentation step aims to  partition  the image into individual and disjointed regions that containing target features (\citealt{2019JPRS..150..115H}), like granules or magnetic features, and the identifying step is to distinguish granules from magnetic features. 
Edge-based techniques (\citealt{j.ins.2003.07.020, Kundu.1986.07.020}) considering edges as boundaries and regions where properties change are applied to segment granules.
Usually, the edges of a feature could be distinguished with high intensity gradients, large intensity discontinuities as well as large intensity variations. 
Granule cells and magnetic features in the photospheric quiet regions are both located in dark granule lanes, and thus the edges could be easily detected. In this work, the segmentation is implemented by finding the edges of these features in granule lanes according to their intensity variations.
A few steps including segmenting and distinguishing granules from magnetic features are described in the following.

\begin{figure}[ht!]\centering
	\includegraphics[width=8cm]{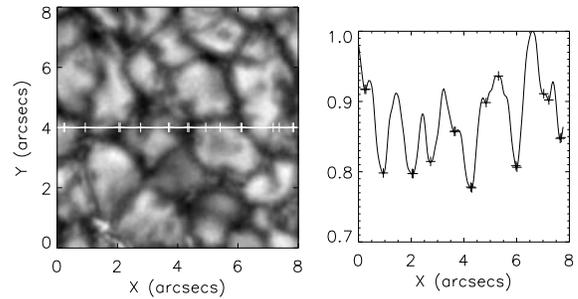}
	\caption{A small image contains granules and granule lanes. Left panel presents a small image that contains a few granules and granule lanes. A dashed line crossed granules and granule lanes, and the white plus signs on the dashed line marked the local intensity valley position in $x$ direction. Right panel gives the intensity variation along the dashed line, and the black plus signs correspond to the positions marked by white plus signs in the left panel.  \label{f01}}
\end{figure}

\begin{figure*}[ht!]\centering
	\includegraphics[width=13cm]{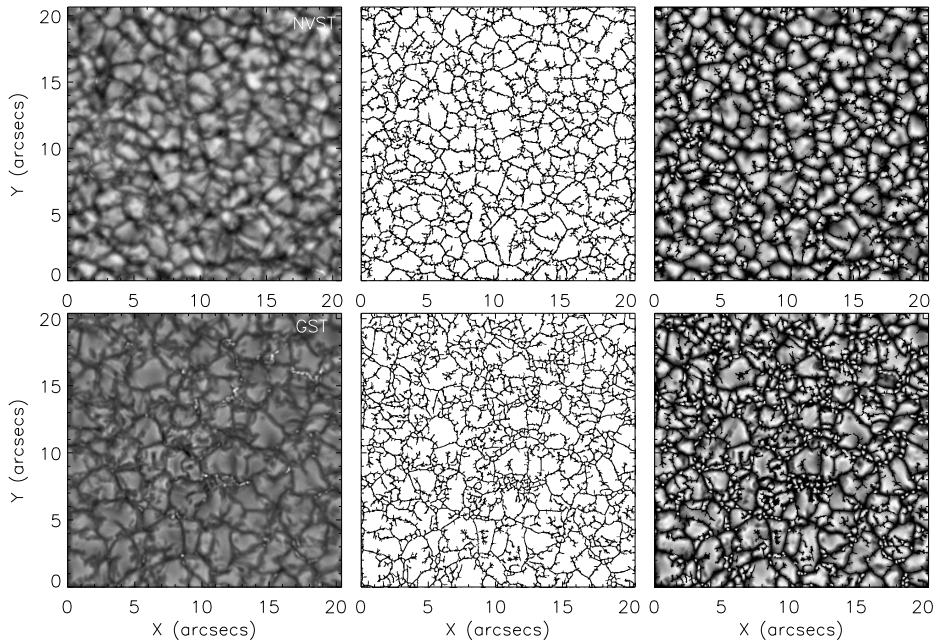}
	\caption{Segment results. The top and bottom rows are the original image, edge detection and the segmenting result from the left to right panels for NVST and GST data, respectively.\label{f02} }
\end{figure*}

(1) Granular edge detection. 
Pixels that possess lowest local intensity values in either $x$, $y$ or diagonal directions are extracted and labelled as edges of granules and magnetic features.
A sample showing the intensity variation along $x$ direction is given in Figure \ref{f01}.
The edge pixels of granules in the left panel marked with white plus signs correspond to local intensity valleys marked with black plus signs in the right panel, which indicates that the edge pixels could be detected by finding the local intensity valleys very well. 
By taking this method, all edge pixels are extracted and the result is given with edges marked in black color in the middle panel of Figure \ref{f02}.

We notice that most of these edge pixels are in granule lanes, and they connect to each other.
A few of them, however, are isolated by these granule lanes and locate on granules. We need to distinguish whether they are real edge pixels or are part of granules, and then discard the false edge pixels. The ``edge pixels'' with intensity higher than ${\mu}$+${\sigma}$ are actually bright pixels on granules, and hereby are false edge pixels. Here the ${\mu}$ and ${\sigma}$ are the mean and standard deviation of intensity of the image. 
The ``edge pixels'' with intensity lower than ${\mu}$-${\sigma}$ are found to locate in granule lanes and considered to be real edge pixels.
However, ``edge pixels'' with intensity in the range of ${\mu}$-${\sigma}$ to ${\mu}$+${\sigma}$ contain both real edge pixels and some false edge pixels.
The false edge pixels in such intensity range locate in the region where granules newly appear and their intensity is originally reduced due to plasma cooling down. 
Such pixels should be discarded since it may cause over-segmentation. A proper size threshold is chosen to take these pixels away, e.g., the edges possessing pixels less than 200 for NVST data are discarded, while the remaining pixels are considered to be real edges of photospheric features. Most of granules are surrounded by these edge pixels and are isolated into individual ones. However, a few of them still connect to each other by sharing a few common pixels. Further steps are needed to disconnect the connecting parts among individual features. 

(2) Separation of closely connecting granules by morphology operations. When two granules stay very close and share common pixels with little intensity variations, part of edges of these two granules overlap, namely common edges, and they are difficult to detect.
Even there are some well-established methods of edge detection, such as Prewitt, Sobel, Marr-Hildreth and Canny detector (\citealt{2017book...216..185L}), these methods mainly search edge pixels by finding the regions where the first-order gradients reach maximum or the second-order gradients change sign.
For example, the Canny detector is good at detecting the edges with obvious local gradients in the direction perpendicular to the edges, but is insensitive to the regions where edge gradients are small.
While, the difficulty of edge detecting in this work is to search for the edge pixels in the regions where 
their intensity fluctuation makes the edge gradients small.
Thus, in such situation edges could not be found very well by detecting either  local intensity valley or large intensity gradients.

We notice that the common edges usually have a few pixels, which could be detected by reducing the boundary pixels of granules a few times.  Boundary pixels could be extracted by taking erode and dilate operations (\citealt{2017book...216..185L}).
Erode operation discards the boundary pixels and fine structures by convolving a specific kernel with target features .
Dilate takes an opposite operation which recoveries the boundary pixels of target features.

For every extracted blob, we need to firstly check whether it contains a single granule or multiple granules, and then make a judgement if we should further segment the granules contained in the blob. 
We firstly apply an erode operation with a kernel of 3${\times}$3 twice to the blob.
If the blob remains an isolated one after both the once and twice erode operations, this blob is considered to contain an individual feature.
If more than one smaller blobs appeared after the once or twice erode operations, the extracted blob is considered to contain multiple features by sharing some common edge pixels. 
For the blobs containing multiple granules, boundary pixels are marked by taking a few times of erode operations and  the overlapped pixels of two granules are searched by taking a few times of dilate operations. The detected common edges are discarded and therefore the close connecting granules inside a blob are segmented successfully.
Till here, the edge detection job is finished and the corresponding segmenting result is presented in the right panel of Figure \ref{f02}.

(3) Distinguishing granule cells from magnetic features.
Figure \ref{f02} shows us that the granules and magnetic features are successfully segmented. We are ready to distinguish granule cells and magnetic features. Magnetic features in quiet Sun are mainly bright points and bright clusters in granule lanes, which are characterized by strong intensity gradient and small sizes. According to the segmenting result in right panel of Figure \ref{f02}, some bright points and bright clusters are separated from granule lanes successfully. Even so, some of them are segmented to be part of granules.
 In order to extract pure granule cells, bright points are firstly identified with the method described in \cite{2018ApJ...856...17L} and then the blobs containing bright points are removed from the segment result.

\section{Results and discussions}\label{sec:Result}
\begin{figure*}[ht!]\centering
	\includegraphics[width=13cm]{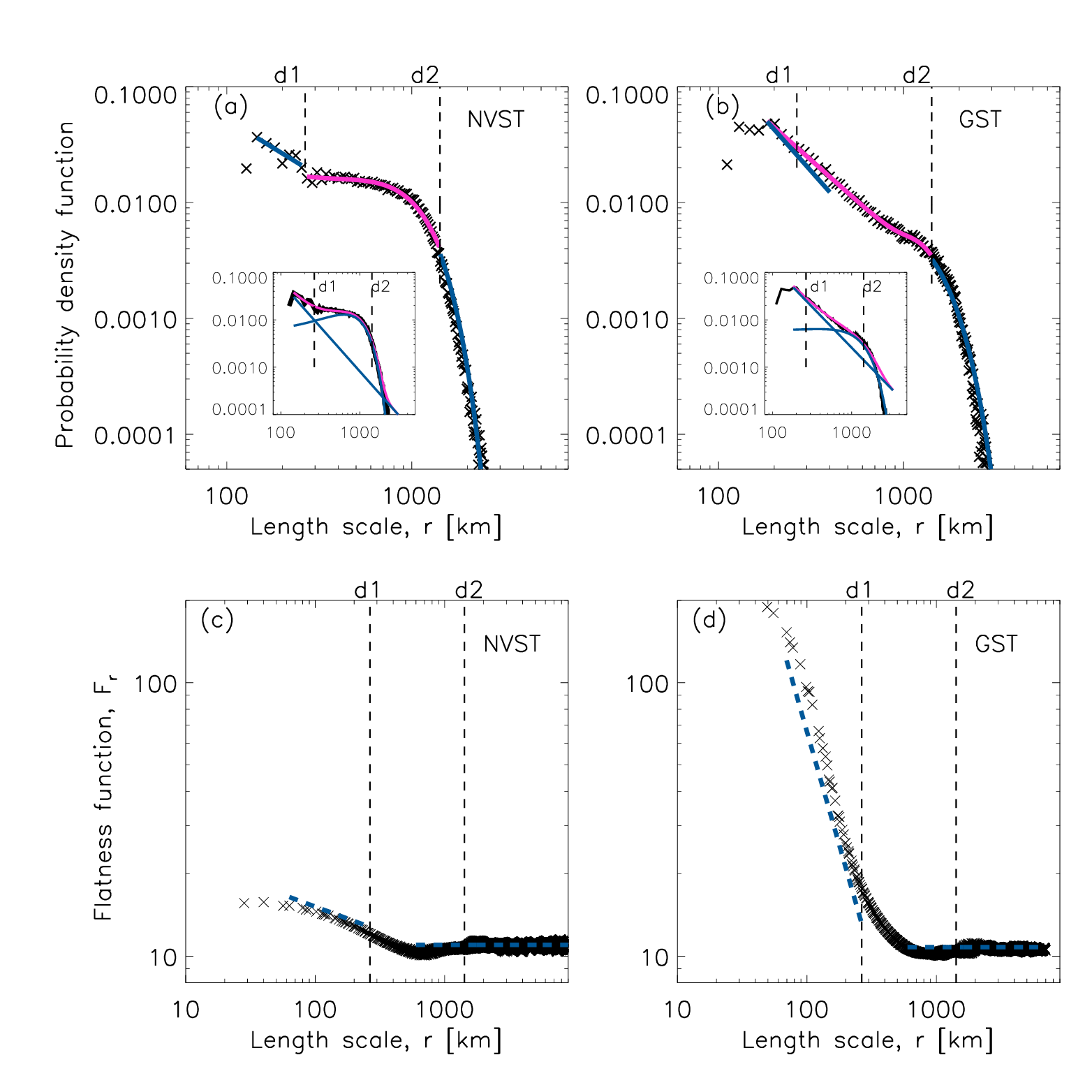}
		\caption{Histogram and flatness functions of granules. Panels (a) and (b) plot the length scale distributions of granules with NVST data and GST data, respectively. Two critical size points are marked in vertical dashed lines at d1=265 km and d2=1420 km. The blue  and red real lines are fitting lines. The data are fitted by power law functions in range of 130 to 265 km for NVST, 170 to 265 km for GST, are fitted by the sum of power law and Gauss functions in the range of 265 to 1420 km, and are fitted by Gauss function in the range of 1420 to 3000 km, respectively. The subplots in panels (a) and (b) are the distributions in black color, fitting results for the data as a whole in red color and the components of fitting functions in blue color, respectively. The critical points are marked at the points where fitting functions deviate the distributions. Panels (c) and (d) are logarithm plots of flatness function with NVST data and GST data. The horizontal blue dashed lines label the ``flat'' length scales of granules, and the slope dashed lines in blue mark the linearity length scales of granules.  \label{f03}}
\end{figure*}

\begin{figure*}[ht!]\centering
	\includegraphics[width=13cm]{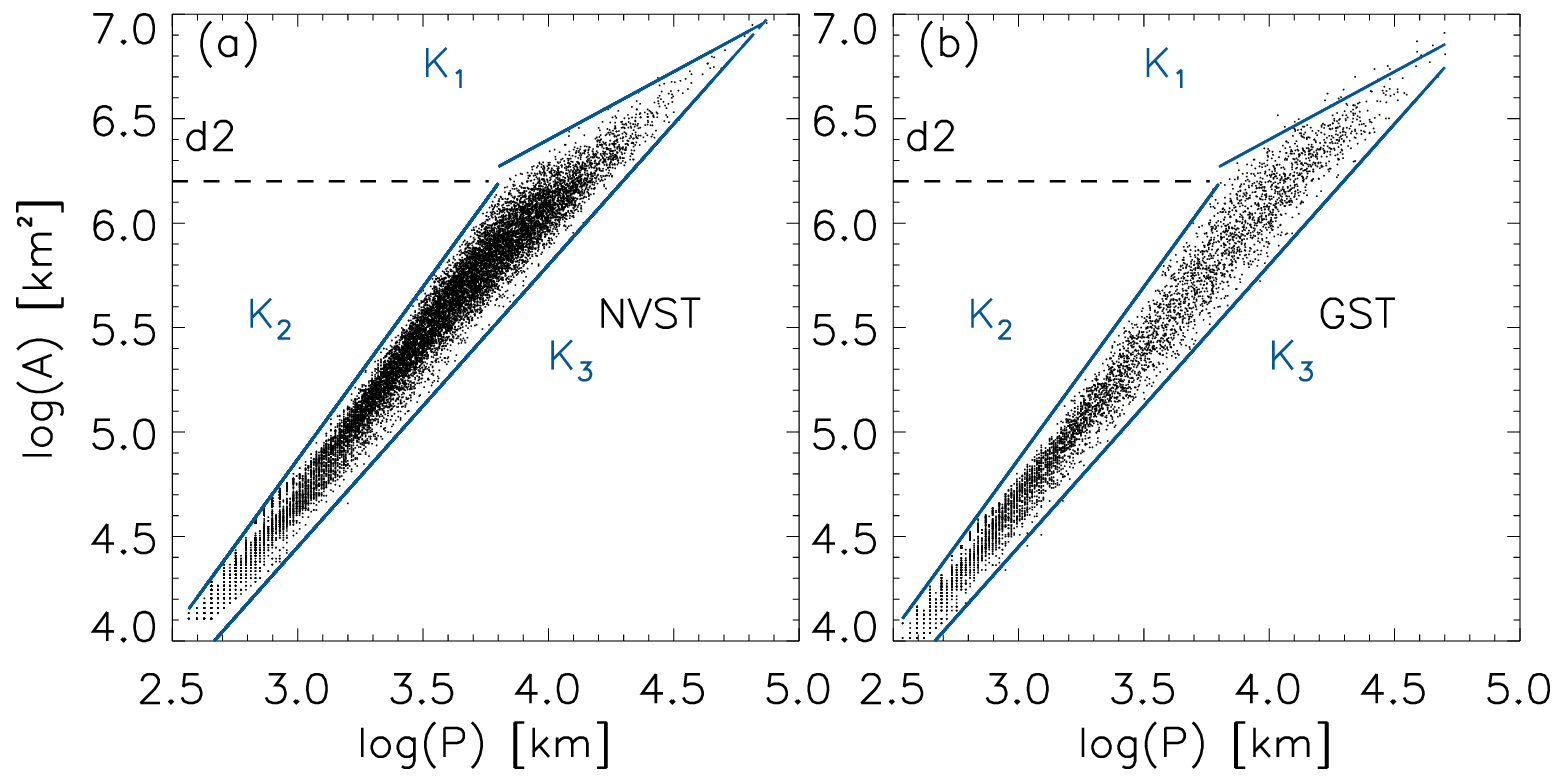}
	\caption{ The P-A logarithm scatter plot. The left and right panels are plotted with NVST data and GST data, respectively.  The slopes are marked with K$_{1}$, K$_{2}$ and K$_{3}$ of 0.65, 1.65 and 1.35, respectively, and the critical size point at 1420 km is labelled at the place where the P-A logarithm plot linearly changes. \label{f04}}
\end{figure*}

The selected NVST data including a sequence of 48 images and GST data including 
a sequence of 77 images have been analyzed. A large number of granules have been successfully extracted with the novel segmenting method. The segmenting method has the ability of detecting both small faint and large bright granules owing to the edge detection method, which is developed to find the local lowest intensity values in specific directions and is independent of intensity threshold. This makes the faint granules accessible, and based on which one can study the granules in the full range of length scales.
The dynamics of granules are analyzed with the distributions of length scale, perimeter and flatness in Figures \ref{f03} and \ref{f04}.

The length of a granule is measured by calculating its equivalent diameter considering the granule as a round shape. Panels (a) and (b) of Figure \ref{f03} present the length scale distribution of granules obtained by NVST data and GST data,  which are
 fitted with different functions.  For the plot in panel (a), the size in the range of 130 to 265 km is fitted by power law function with slope of -0.9; the size in the range of 265 to 1420 km is fitted by the sum of a power law and a Gauss function in red solid line; and the size in the range of 1420 - 3000 km is fitted by Gauss function. The subplot draws the power law and Gauss functions in blue color as well as their sum in black color. For the plot in panel (b), the size in the range of 170 to 265 km is fitted by power law function with slope of -1.8; the size in the range of 265 to 1420 km is fitted by the sum of a power law and a Gauss function in red solid line; and the size in the range of 1420 - 3000 km is fitted by a Gauss function. The subplot shows their fitting functions as well.

 The power law and Gauss fitting functions are expressed in equations \ref{equ0} and \ref{equ1}:
\begin{equation}\label{equ0}
	Y_{1}=C_{1}x^{k} \\
\end{equation}

\begin{equation}\label{equ1}
	Y_{2}=C_{2}\frac{1}{\sqrt{2{\pi}}{\sigma}}e^{(-\frac{1}{2}(\frac{x-{\mu}}{{\sigma}})^{2})} \\
\end{equation}
The goodness of fitting is also measured by calculating the Chi-Square (${\chi}^2$) with the formula \ref{equ2}:
\begin{equation}\label{equ2}
	{\chi}^2=\sum{\frac{(Y_{i}-{Y_{fit}}_{i})^{2}}{Y_{i}}}, \\
\end{equation}
where $Y_{i}$ and $Y_{fit_{i}}$ are the statistical and theoretical values.
The fitting parameters and the measured Chi-Square (${\chi}^2$) are listed in Table \ref{para2}.

The two critical size points at d1= 265 km and d2= 1420 km have been found and marked in both panels (a) and (b) with vertical dashed lines.
They are obtained at the places where fitting functions change.
According to panels (a) and (b) of Figure \ref{f03} these two critical size points might separate the motions of granules into three types: the convection, the mixing motions of convection and turbulence, and turbulence. Furthermore,  panel (b) of Figure \ref{f03} demonstrates almost linearly, and power law function plays more and more important role along with the length scale reducing in the mixing motion stage.
	This reveals that the motions of granules  become obviously turbulent with the decrease of length.

We notice that the distribution shapes in panels (a) and (b) are obviously different. The length scale range from 265 km to 1420 km in panel (b) behaves more linearly than that in panel (a).
	In such length scale range, granules in panel (b) of Figure \ref{f03} demonstrates more turbulent than that in panel (a). 
	Such obvious discrepancy is probably caused by different spatial resolutions since the GST data possess higher spatial resolution which could provide more detail information of intensity variations of granules. On the other hand, there are some granules with sizes smaller than 130 km in panel (a) and smaller than 170 km in panel (b) drops off the linear fitting lines, which occupies a quite high probability. This phenomenon is also reported by \cite{1986SoPh..107...11R, 1997ApJ...480..406H, 2012ApJ...756L..27A}, and is considered being independent of granule segmenting methods. It is probably caused by the limited spatial resolutions which might underestimate the population of tiny granules. While the details of such tiny granules are beyond the scope of this work and will be analyzed in future, hopefully with even higher spatial resolution telescope, such as Inouye Solar Telescope.

In addition, it is necessary to note here that the size threshold applied in granule segmenting steps is different from the derived critical sizes.
The size threshold in the granule segmenting algorithm is used to distinguish if the edge pixels are false edges or not. Usually the false pixels are part of granules, which are isolated by pixels in granule lanes when put all these detected edge pixels into an empty array and set them value of 1. A proper threshold could help discard the false edge pixels from all detected edge pixels. One obvious difference between the critical sizes and threshold size is that the former is value of granule size, and the latter is part of granules which might develop into the edges of granules.

\begin{deluxetable*}{ccccccccc}
	\tablenum{1}
	\tablecaption{Fitting parameters of diameter and area distribution \label{para2}}
	\tablewidth{0pt}
	\tablehead{
\colhead{} &	\colhead{Length} & \colhead{$k$}&\colhead{$C_{1}$}&\colhead{$C_{2}$}&\colhead{Mean}&\colhead{Standard divation}&\colhead{Chi-Square} &\colhead{Fitting}\\
\colhead{} &\colhead{ranges [km]} & \colhead{}&\colhead{}&\colhead{}&\colhead{${\mu}$}&\colhead{${\sigma}$}&\colhead{ (${\chi}^{2}$)} &\colhead{functions}}
	\startdata
&130-265    &-0.98 & 4.8 & -        &-    &-  &0.002&$Y_{1}$\\
NVST&265-1420    &-1.4  &  14 & 19&517  &548&0.002&$Y_{1}$+$Y_{2}$\\
&1420-3000    & -    &   -        & 19& 517 &548&0.004&$Y_{2}$  \\
\hline
\hline
&170-265    &  -1.8   & 363 &-&-&-&0.001&$Y_{1}$ \\
GST&265-1420    &  -1.8   & 363 &15&629&467&0.001 &$Y_{1}$+$Y_{2}$\\
&1420-3000    &  -   & - &15&629&467&0.001 &$Y_{1}$\\
	\enddata
\end{deluxetable*}

We also calculated the flatness function, $F(r)$, in this work to confirm the critical points obtained in the length scale distribution plots.
According to \cite{2012ApJ...756L..27A}, sixth order flatness function describes the ratio between the sixth order structure function to the cube of second order structure function (\citealt{2005SoPh..228...29A,2012ApJ...756L..27A,2010ApJ...722..122A}), where the $qth$ order structure function $S_{q}(r)$ could be written as $qth$ power of the intensity increment between two pixels at $(x+r)$ and $x$ in distance $r$:
\begin{equation}\label{equ4}
S_{q}(r)=|I(x+r)-I(x)|^{q}
\end{equation}
The sixth order flatness function, $F(r)$, hereby, could be expressed as:
\begin{equation}\label{equ5}
	F(r)=\frac{<|I(x+r)-I(x)|^{6}>}{(<|I(x+r)-I(x)|^{2}>)^{3}}, \\
\end{equation}
where $<>$ stands for the mean value of the structure function in the whole image. 
The flatness could be used to measure the intermittency and multifractality of features. 
Intermittency and multifractality are the typical characteristics of turbulence.
The linearity of flatness with negative slope in logarithm plot reveals the multifractality and intermittency of features;
while the ``flat'' shape of flatness which is derived by Gaussian distribution, indicates none intermittency and none multifractality of features (\citealt{2005SoPh..228...29A, 2010ApJ...722..122A}).

The panels (c) and (d) of Figure \ref{f03} present the flatness functions obtained with NVST and GST data, respectively, which demonstrate in linearity, nonlinearity, and ``flatness''  in three length scale ranges separated by two critical points at 265 km and 1420 km.
The flatness functions with sizes larger than critical point of 1420 km is ``flat'', which indicate the granules of being non-intermittent, non-multifractal and regular, and they correspond to the Gaussian distribution presented in panels (a) and (b) of Figure \ref{f03}; 
the flatness functions demonstrate nonlinearity with sizes in the range of 265 to 1420 km, which correspond to the mixing motions of convection and turbulence as revealed by the sum of Gauss and power law fitting functions in the same length range in panels (a) and (b) of Figure \ref{f03};  
and the flatness functions show linearity with negative slopes with sizes smaller than 265 km, which reveals the granule motions of being fully turbulent as described by power law distributions in the panels (a) and (b) of Figure \ref{f03}.

The flatness function presented in panel (d) shows more details than that in panel (c) due to higher spatial resolution of GST. Especially
in the nonlinearity length scale range, panel (d) shows  $F(r)$ value vastly increases along with the granule size reducing. Correspondingly,  the distribution in panel (b) of Figure \ref{f03} in same length range demonstrates nearly linear. This reveals that the process of transferring dynamics from convection to turbulence is dominant and significant.

\cite{2012ApJ...756L..27A} also analyzed the GST data and reported two populations of granules: the regular granules with size larger than 1000 km are convective;  the mini granules with sizes smaller than 600 km are intermittent and turbulent. 
{In general, this is in agreement with the result presented in panel (b) of Figure \ref{f03} in this work since the length scale distribution of granules in mixing motion stage behaves strongly turbulent. The difference is that we classify this length range scale into mixing motion stage which combines the convection and turbulence motions. On the other hand, the critical point of 1420 km which separate larger granules as regular is obtained at  the point where fitting function changes in this work, while the critical point of 1000 km is derived by the mean value of Gauss fitting function in \cite{2012ApJ...756L..27A}.}
We also notice that previous studies separated the motions of granules into convection and turbulence by a single critical point (\cite{1990VA.....33..413G, 1986SoPh..107...11R, 2012ApJ...756L..27A}). The reliable granule segmenting method applied in this work and the high-quality data obtained by large aperture solar telescopes enable us to detect more details of granules, and hereby find the three regimes of motions, namely, convection, mixing motion of convection and turbulence, and turbulence.
In these three regimes, the length scale distribution of granules follows Gauss function, sum of Gauss and  power law functions, power law, separately, and the corresponding flatness function demonstrates ``flatness'', nonlinearity, and linearity, respectively.  

Additionally, the fractal dimension, $D$, which could be used to measure the shape irregularity, is also  calculated in this work by taking the following  the area-perimeter relation
\begin{equation}\label{equ3}
	P=K{\cdot}A^{D/2} \\
\end{equation}
where $P$, $K$, $A$ are perimeter, factor and area, respectively. 
The $P$-$A$ log-log scatter plots are presented in panels (a) and (b) of Figure \ref{f04} with NVST data and GST data, respectively, which demonstrate a linear relationship with a sharp change at 1420 km. The slopes are found to vary in the ranges of [1.35, 1.65] and [0.65, 1.35], and the corresponding fractal dimensions $D$ of [1.21, 1.48] and [1.48, 3.07] with sizes smaller and larger than 1420 km, respectively. This might indicate that the size at 1420 km is the separator between regular granules and turbulent granules. 
\cite{1986SoPh..107...11R}  found  fractal dimension $D$ of 1.25 for granules smaller than 900 km (1.37{\arcsec}). According to the Kolmogorov theory of predicting $D$ of $5/3$ and $4/3$ for isotherms and isobars, respectively, they found that $D=1.25$ is close to $4/3$ and hereby suggested that the small scale granules (smaller than 1.37{\arcsec}) are turbulent origin.
\cite{1990VA.....33..413G} found $D$ of 1.13 and 1.9 for granules smaller and larger than the critical size of 1.32{\arcsec}, respectively, 
and they argued that the fractal dimension might not be correct for small granules since the shapes of small granules are roughly considered as round because of limited spatial resolution.


\section{Conclusion}\label{sec:Conclusion}
In this work, we studied the dynamical motions of photospheric granules with the high spatial resolution data from NVST and GST. To this end, we developed a new granule segmenting method, which enables us to identify the small faint granules located in granule lanes as well as large bright granules, thus allowing us to access the full length range of granules.

The length scale distributions of granules obtained with NVST data and GST data have been studied, and two critical points at sizes of d1=265 km and d2=1420 km have been found. The critical point at size of 265 km is determined at the cross point of the Gauss function and Power law function for NVST data and at the point where the linear fitting deviate from the data for GST data; the critical point at size of 1420 km is determined at the point where the Gauss function starts to fit the length scale distribution while the sum of Gauss function and Power law function starts to deviate.
The granules with size larger than 1420 km can be fitted by Gauss function and are thought to be dominated by convective motions. The granules with sizes in the range of 265 - 1420 km can be fitted by the sum of Gauss function and Power law function, which indicates a mixing motion of the convection and turbulence. Even though in the mixing motion stage, the granules behave strongly turbulent as revealed by GST data. Granules with size smaller than 265 km can be fitted by power law function, suggesting that the ``mini'' granules are fully turbulent. 
Correspondingly, the flatness functions plotted by both the NVST and GST data are ``flat'', linear with negative slope, and non-linear with granule sizes above 1420 km, below 265 km, and between  265 and 1420 km, respectively.
In addition, the fractal dimension $D$ has been analyzed through the $A$-$P$ logarithm plot, and the critical point at 1420 km which separate the granule into the turbulence and stable convection.

 In a summary, our analysis suggests that a cascade process occurs in the dynamical motions of granules: the large granules or regular granules are originally convective. They split and transport energy into small granules. Turbulent motions generate, start to grow and compete with convective motions. The small granules continue to split into smaller ones. The turbulence becomes strong enough, overcomes the convective motion and dominates the motions at last. 
\\

This work is supported by the National Natural Science Foundation of China (Nos. 41822404, 11973083, 42074205, 11763004, 11729301, 11803005, 12111530078, 1217030170, 4217040250 ) and Shenzhen Technology Project ( JCYJ20190806142609035, GXWD20201230155427003-20200804151658001).
We are grateful for the topic observation chance provided by Fuxian Solar Observatory (FSO), and for the assistance of observations and data reconstructions provided by Professor Xu Zhi, observation assistants Luo Lin and Liao Guangquan, and the NVST team. We also thank the GST team for providing high quality data. BBSO operation is supported by NJIT and US NSF AGS-1821294 grant. GST operation is partly supported by the Korea Astronomy and Space Science Institute, the Seoul National University, and the Key Laboratory of Solar Activities of Chinese Academy of Sciences (CAS) and the Operation, Maintenance and Upgrading Fund of CAS for Astronomical Telescopes and Facility Instruments.

\bibliography{sample63}{}
\bibliographystyle{aasjournal}



\end{document}